\documentclass[%
reprint,
superscriptaddress,
 amsmath,amssymb,
 aps,
]{revtex4-2}

\usepackage{graphicx}%
\usepackage{multirow}%
\usepackage{amsmath,amssymb,amsfonts}%
\usepackage{amsthm}%
\usepackage{mathrsfs}%
\usepackage[title]{appendix}%
\usepackage{xcolor}%
\usepackage{textcomp}%
\usepackage{booktabs}%
\usepackage{listings}%
\usepackage{indentfirst}
\usepackage{float}
\usepackage{hyperref}
\hypersetup{
    colorlinks=true,
    linkcolor=blue,
    filecolor=blue,      
    urlcolor=blue,
    citecolor=blue
    }

\newcommand{\ket}[1]{\left|{#1}\right\rangle}

\usepackage{titlesec}
\titleformat*{\section}{\centering\footnotesize\bfseries\uppercase}

\usepackage[paperwidth=210mm,paperheight=297mm,centering,hmargin=1.7cm,vmargin=1.5cm]{geometry}

\begin{document}

\title{\large \textbf{In-situ Imaging of a Single-Atom Wave Packet in Continuous Space}}

\author{Joris Verstraten}
\affiliation{Laboratoire Kastler Brossel, ENS-Universit\'{e} PSL, CNRS, Sorbonne Universit\'{e}, Coll\`{e}ge de France, 24 rue Lhomond, 75005, Paris, France}
\author{Kunlun Dai}
\affiliation{Laboratoire Kastler Brossel, ENS-Universit\'{e} PSL, CNRS, Sorbonne Universit\'{e}, Coll\`{e}ge de France, 24 rue Lhomond, 75005, Paris, France}
\author{Maxime Dixmerias}
\affiliation{Laboratoire Kastler Brossel, ENS-Universit\'{e} PSL, CNRS, Sorbonne Universit\'{e}, Coll\`{e}ge de France, 24 rue Lhomond, 75005, Paris, France}
\author{Bruno Peaudecerf}
\affiliation{Laboratoire Collisions Agr\'egats R\'eactivit\'e, UMR 5589, FERMI, UT3, Universit\'e de Toulouse, CNRS, 118 Route de Narbonne, 31062, Toulouse CEDEX 09, France}
\author{Tim de Jongh}
\affiliation{Laboratoire Kastler Brossel, ENS-Universit\'{e} PSL, CNRS, Sorbonne Universit\'{e}, Coll\`{e}ge de France, 24 rue Lhomond, 75005, Paris, France}
\author{Tarik Yefsah}
\affiliation{Laboratoire Kastler Brossel, ENS-Universit\'{e} PSL, CNRS, Sorbonne Universit\'{e}, Coll\`{e}ge de France, 24 rue Lhomond, 75005, Paris, France}

\begin{abstract}
 \quad The wave nature of matter remains one of the most striking aspects of quantum mechanics. Since its inception, a wealth of experiments has demonstrated the interference, diffraction or scattering of massive particles. More recently, experiments with ever increasing control and resolution have allowed imaging the wavefunction of individual atoms. Here, we use quantum gas microscopy to image the in-situ spatial distribution of deterministically prepared single-atom wave packets as they expand in a plane. We achieve this by controllably projecting the expanding wavefunction onto the sites of a deep optical lattice and subsequently performing single-atom imaging. The protocol established here for imaging extended wave packets via quantum gas microscopy is readily applicable to the wavefunction of interacting many-body systems in continuous space, promising a direct access to their microscopic properties, including spatial correlation functions up to high order and large distances.
\end{abstract}

\maketitle

In 1924, Louis de Broglie proposed his theory of electronic waves, where he merged the concepts of particles and waves through the notion of wave-particle duality, which became foundational to the theory of quantum mechanics. A massive particle is associated with a wavefunction whose dynamics is governed by the Schr{\"o}dinger equation. This was confirmed in the following years in a number of seminal experiments, such as the celebrated observation of electron diffraction by Davisson and Germer \cite{davisson1928}, and the diffraction of helium atoms off a crystal surface by Estermann and Stern \cite{estermann1930}, demonstrating the wave nature of composite particles. Since these historic measurements, wave behavior of massive particles has been evidenced in a wide range of experiments: wave packets were revealed through interference, diffraction or scattering \cite{egerton2005, ashfold2006,cronin2009,stodolna2013,rauch2015,mcdonald2016,perreault2017,waitz2017,guan2019,karamatskos2019, heazlewood2021,paliwal2021,deJongh2022a}, and for particles of increasing size and complexity, from elementary particles such as electrons \cite{hawkes1996, tonomura1989} to composite systems such as atoms, clusters, and molecules composed of up to 2,000 atoms \cite{fein2019,hornberger2012}.

Another class of experiments has allowed probing the spatial distribution of individual wave packets. For instance, in quantum dots, the static spatial probability density of single electrons has been measured directly or indirectly via scanning tunneling spectroscopy \cite{vdovin2000,maltezopoulos2003}. In hybrid atom-ion systems, the vibrational wave packet dynamics of single molecules within their molecular potential was directly observed using high-resolution ion microscopy \cite{zou2023}. On ultracold atom platforms, the tunneling dynamics of single-particle wave packets in a periodic potential were observed with various imaging methods \cite{karski2009, weitenberg2011, fukuhara2013, genske2013, preiss2015, young2022}. In atom tweezer harmonic traps, the squared modulus of the wavefunction in momentum space has been probed for few-particle \cite{holten2021} and single-particle states \cite{brown2023a}.

\begin{figure}[!t]
    \centering
    \includegraphics[width=0.95\linewidth]{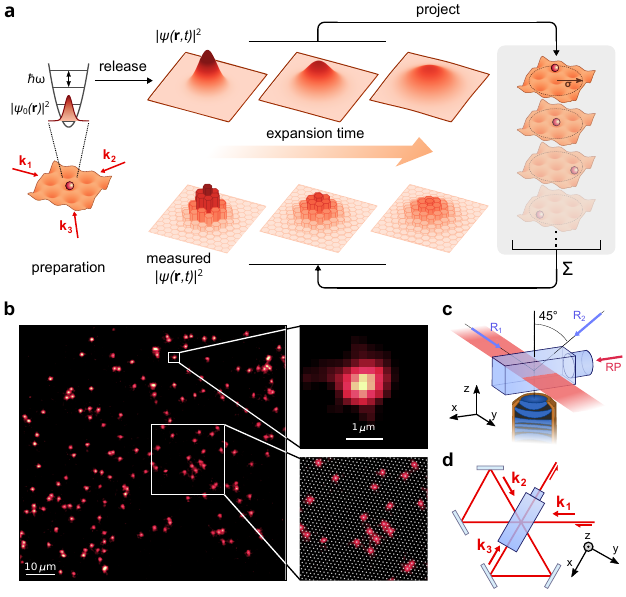}
    \caption{\textbf{Preparation and \textit{in-situ} imaging of single-atom wave packets.} \textbf{(a)} Measurement scheme: Individual atoms are prepared close to the harmonic oscillator ground state of individual sites of a triangular optical lattice created by a self-interfering laser beam with wave vectors $\mathbf{k_1,k_2}$ and $\mathbf{k_3}$. Wave packets initially trapped in the lattice wells, characterized by a Gaussian probability density distribution $|\psi_0(\mathbf{r})|^2$, are released in a plane, allowing them to expand for a given time. For imaging after expansion, the lattice is quickly ramped up again, projecting the wave packet, and Raman sideband cooling is applied to pin the atom on a single site. Resulting atomic positions are recorded through site-resolved fluorescence imaging. From many repetitions of identically prepared wave packets we create histograms of the projected positions with a discretization given by the lattice structure, resulting in a measured probability distribution $|\psi(\mathbf{r},t)|^2$. \textbf{(b)} Experimental single-atom resolved image. The top-right panel shows a subregion containing an individual atom. The bottom-right panel displays an enlarged region of the image over which the reconstructed triangular lattice structure with a spacing of $a_L = 709$\,nm is shown as white dots. \textbf{(c)} Experimental configuration of the oblate optical dipole trap confining the atoms to a two-dimensional plane, the Raman beams (R$_1$, R$_2$ and RP) used for cooling and imaging, and the microscope objective. \textbf{(d)} Top view of the experimental configuration, showing the geometry of the optical lattice beam.}
    \label{fig:general_procedure}
\end{figure}

\begin{figure*}[!t]
    \centering
    \includegraphics[width=0.97\textwidth]{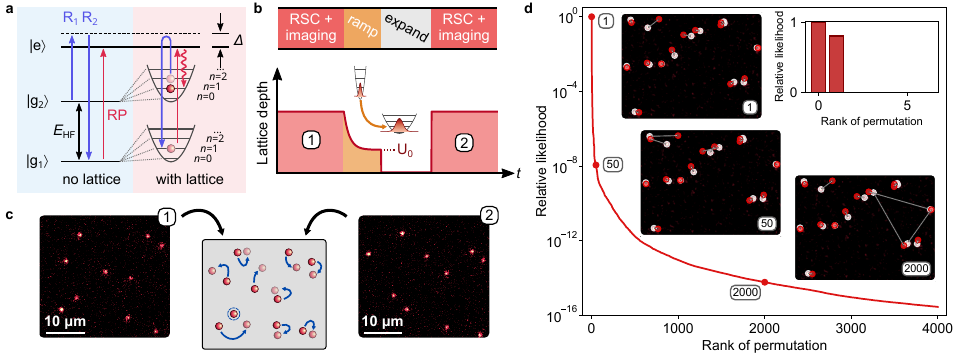}
    \caption{\textbf{Experimental sequence and image analysis.} \textbf{(a)} Raman sideband cooling scheme for $^6$Li atoms. The two Raman beams ($R_1$ and $R_2$) are blue-detuned by $\Delta = h\cdot 3$\,GHz with respect to the first electronically excited state $|e\rangle$ and connect the two hyperfine levels $|g_1\rangle$ and $|g_2\rangle$ in the ground state manifold, split by $E_\mathrm{HF} = h\cdot 228.2$\,MHz. A resonant repumper (RP) beam connects $|g_1\rangle$ and $|e\rangle$. In the presence of a deep optical lattice, RSC brings the atoms to lower harmonic oscillator eigenstates, indicated by eigenvalue $n$, while producing fluorescence photons for single-particle imaging through the microscope objective, serving as both a preparation and detection method. \textbf{(b)} Experimental sequence for the preparation, expansion and pinning of the single-atom wave packets. After the wave packets have been localized in a deep optical lattice and their initial positions recorded in a first image (1), RSC is turned off and the lattice depth is adiabatically ramped down to a variable value $U_0$ to adjust the width of their initial momentum distribution. The lattice is then suddenly switched off and the wave packets expand for a time $t$ after which we take a second image (2) to record the new atom positions. \textbf{(c)} Two single atom images taken in a single experimental realization (left and right panels). The center panel schematically shows the most likely assignment of how the atoms moved between the two images. \textbf{(d)} Relative likelihood of the 4000 most likely assignments based on a combined likelihood computation, for $\omega = 2\pi \times 600(30)\,\mathrm{kHz}$ ($U_0 = 0.38\,U_\mathrm{max}$) and an expansion time of $10\,\mu$s. The image insets show the assignments ranked 1$^\mathrm{st}$, 50$^\mathrm{th}$ and 2000$^\mathrm{th}$. Initial (white dots) and assigned final (red dots) positions are connected by arrows. The top-right inset shows the relative likelihood in a linear scale.}
    \label{fig:sequence}
\end{figure*}

Here, we prepare single atoms near the ground state of harmonic oscillator wells and probe the associated Gaussian wave packets in-situ by following their dynamics upon release from the trap as they expand in a plane. By varying the initial momentum spread of the single-atom wave packets, we observe dynamics that is in quantitative agreement with the prediction from the Schr{\"o}dinger equation. Our measurement represents a pristine observation of the textbook ballistic expansion of a single-atom Gaussian wave packet in real space. This is realized by performing quantum gas microscopy \cite{bakr2009,sherson2010,gross2017,gross2021} after projecting the wavefunction freely expanding in space onto a deep optical lattice \cite{bakr2010,brown2020,buob2023}. Using the known single-particle wave packet expansion, we quantitatively determine a protocol for controlled projection of a wavefunction evolving in continuous space and reliable pinning of the corresponding atom for imaging. With this, we achieve a crucial pre-requisite to extend the use of quantum gas microscopy to interacting many-body systems in continuous space, offering direct access to spatially-resolved correlation functions up to high order and at large distances.

\section*{Preparing single-atom wave packets}\label{sec:experimental_procedure}

\begin{figure*}[!ht]
    \centering
    \includegraphics[width=0.95\textwidth]{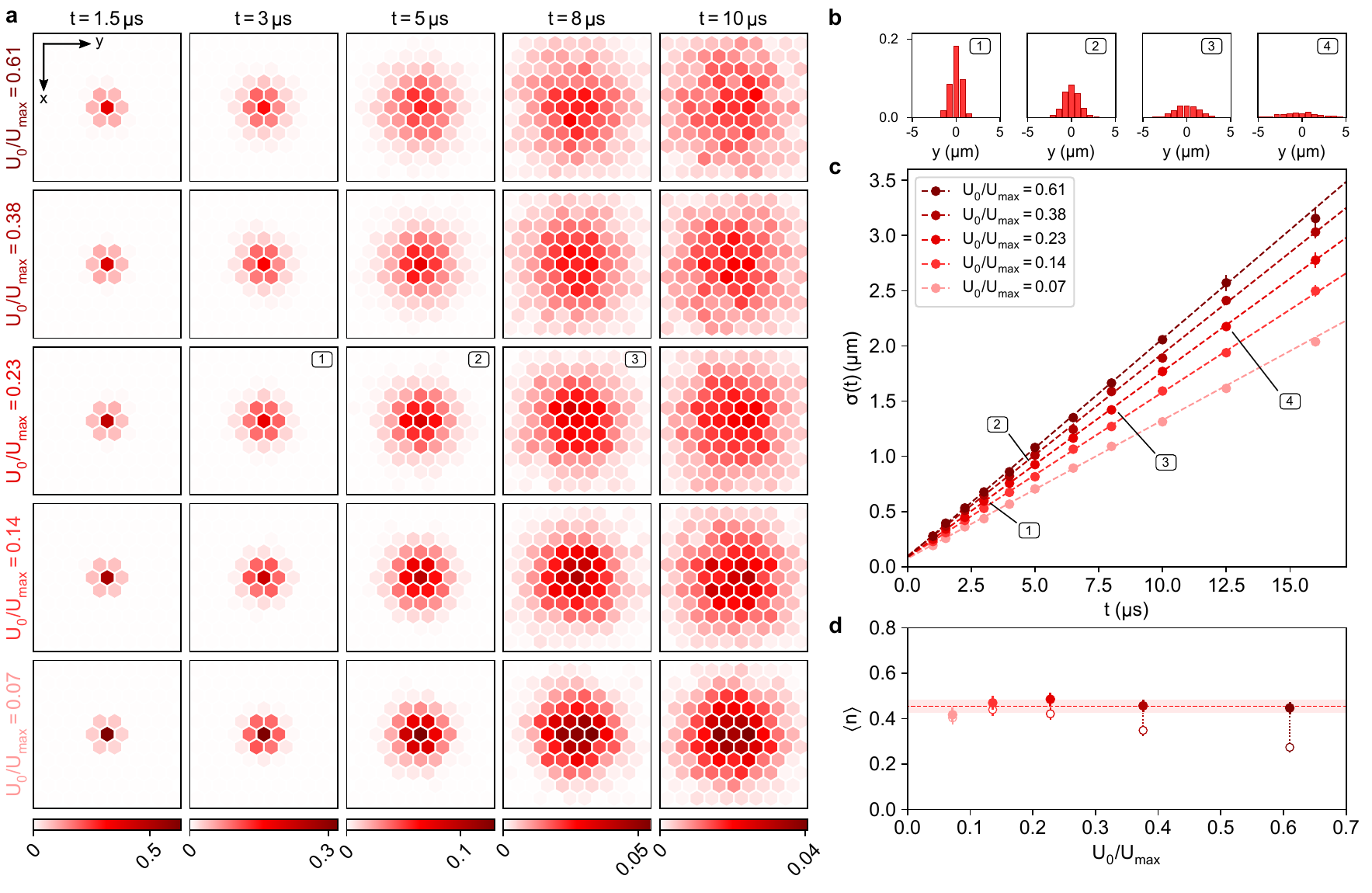}
    \caption{\textbf{Expansion dynamics of a single-particle Gaussian wave packet.} \textbf{(a)} Probability density distribution as a function of expansion time $t$ for wave packets prepared at different lattice depths $U_0$ before release. Hexagons represent the lattice sites of the triangular pinning lattice, with the central site indicating the position of each atom in the first single-atom image. The histograms are obtained after 100 measurements, each with around $20$ to $50$ identically prepared single-atom wave packets. \textbf{(b)} Cuts of the two-dimensional histograms at $U_0 = 0.23\,U_\mathrm{max}$ for $t=3\,\mu$s, $5\,\mu$s, $8\,\mu$s, and $12.5\,\mu$s. \textbf{(c)} Data points show the width of the Gaussian wave packet ($\sigma$) extracted from the probability density distribution as a function of $t$ for each preparation lattice depth. Error bars indicate the standard error of the maximum likelihood estimate and are typically smaller than the symbol size. Dashed lines represent linear fits to the experimental data. \textbf{(d)} Solid points show the average harmonic oscillator eigenvalues $\langle n \rangle =[0.42(3),0.47(3),0.49(3),0.46(3),0.45(3)]$ for each preparation, corresponding to the lattice depth $U_0/U_\mathrm{max} = 0.072(5), 0.136(5), 0.228(7), 0.376(10)$ and $ 0.61(2)$. The open points are the values extracted from the linear fits in (b) using Eq.\:(\ref{eq:expand}) before the correction due the non-instantaneous release is taken into account (see text). The horizontal red dashed line together with the light red band indicate the average of all five values $\langle n \rangle=0.46(3)$.}
    \label{fig:mosaic}
\end{figure*}

We start with a dilute ensemble of Lithium 6 ($^6$Li) atoms prepared near the ground state of the wells of a deep optical lattice. The optical lattice is generated by the self-interference of a laser beam in the horizontal $x-y$ plane forming three arms crossing at 120-degree angles (Fig.\:\ref{fig:general_procedure}a), with a vertical polarization that results in a triangular lattice geometry \cite{jin2024}. Close to the bottom of the wells, the potential seen by the atoms is well approximated \footnote{For the sake of clarity we here ignore a slight difference between the frequencies in the $x$ and $y$ directions. This is however taken into account in the data analysis.} by a harmonic oscillator $U(x,y) \simeq \frac{1}{2} m \omega^2 \left(x^2 + y^2 \right)$, with $\omega\simeq2\pi\times1\,\mathrm{MHz}$, leading to eigenstates $\ket{n_x,n_y}$ and associated energy $\left(n_x +n_y + 1 \right) \hbar\omega$. In the vertical direction ($z$-axis), the atoms are confined in a single plane using a laser-light sheet \footnote{Additional details on the experimental setup and the data analysis can be found in the supplementary materials}. We typically load a few tens of atoms over the $\sim6000$ available lattice sites in order to sparsely populate the lattice. The inter-particle spacing is on the order of $10\,\mu$m, more than ten times the lattice spacing $a_{\rm L}=709$\,nm, such that each occupied well contains no more than one atom. The loading of tens of independent atoms allows us to obtain as many realizations of the single-particle expansion dynamics in a single experimental run.

In order to bring the atoms near the ground state of the lattice wells, we apply Raman sideband cooling (RSC) \cite{Note2}, as depicted in Fig.\:\ref{fig:sequence}a: a two-photon process drives a hyperfine transition $\ket{F=3/2,n_i} \to \ket{F=1/2, n_i-1}$ that reduces the vibrational level $n_i$ (with $i=x,y$) of atoms in each lattice site, while a third near-resonant beam excites the atoms in $\ket{F=1/2, n_i-1}$ to the upper state $\ket{F'=1/2}$, from which they in turn decay to the original hyperfine state $\ket{F=3/2,n_i-1}$. At the end of one RSC cycle, the motional energy of a given atom has been reduced by one quantum. After a number of cycles, and in the absence of heating sources, this would result in preparing the atom in the ground state wavefunction of the harmonic oscillator well. The photons that are spontaneously emitted during the decay from the excited state $\ket{F'=1/2}$ to the ground state $\ket{F=3/2}$ (or $\ket{F=1/2}$), serve as a fluorescence signal to detect each atom while it is held in a given site. This dual-purpose approach of RSC is routinely used in quantum gas microscopy \cite{gross2017,gross2021}, where it was developed for the study of fermionic systems in optical lattices \cite{cheuk2015, parsons2015, omran2015}. Our imaging method is identical to quantum gas microscopy, except that we apply it to probe wave packets in continuous space. In Fig.\:\ref{fig:general_procedure}b, we show a typical single-atom image obtained with our apparatus.

The RSC beam parameters have been optimized to reach the highest fidelity while bringing the atoms as close as possible to the vibrational ground state of the lattice wells. Based on the measured RSC beam and lattice parameters, we expect average vibrational numbers $\left\langle n_x \right\rangle\approx\left\langle n_y\right\rangle\simeq 0.4$ at the end of the cooling procedure \cite{Note2}, which we determine experimentally in the following. In this first step, we therefore not only prepare an ensemble of single-atom Gaussian wavefunctions near the ground state of harmonic oscillators, but also have an image of their initial positions with a resolution at the level of a single lattice site. 
    
\section*{Release, expansion, and detection of wave packets}

The diagram of the experimental sequence is shown in Fig.\:\ref{fig:sequence}b. Following the initialization phase described above, we turn off all RSC laser beams, and adiabatically ramp down the lattice depth to a variable fraction $U_0$ of its initial value $U_{\rm max}$, which allows us to control the frequency of the harmonic oscillator $\omega\propto \sqrt{U_0}$ while maintaining the wave packets in their initial vibrational level. We then suddenly turn off the lattice beams, keeping the light sheet potential on, to let the atomic wave packets expand in the $x-y$ plane for a variable time $t$. We subsequently turn the lattice back on to its maximal depth using an optimal ramp time (see below), and perform RSC again to pin the atoms. This step projects the free atoms to the nearest lattice site after expansion. We thus obtain a second image displaying the final positions of the atoms (see Fig.\:\ref{fig:sequence}c). Assuming no bias is introduced upon pinning from continuous space and imaging, other than a discretization of space, each atom on the second image will be detected at a certain position, sampling the probability density - i.e., the modulus squared - of the wavefunction after release. By relating each atom of the second image to its original lattice site in the first image from many realizations \cite{Note2}, we obtain histograms in position space representing the probability density of the expanding wave packets at variable time $t$. 

In order to reliably assign each atom detected after expansion to its initial position, we use a self-consistent analysis based on a maximum likelihood estimate, which we describe in detail in the supplementary materials \cite{Note2}. Finding the permutation of atoms that maximizes the total likelihood reduces to the linear assignment problem. In Fig.\:\ref{fig:sequence}d, we show an example of relative likelihood as a function of permutation rank, showing a rapid decrease of the probability. The experimental parameters were chosen such that the typical distance travelled by each atom is small compared to the interparticle distance, which strongly facilitates the analysis as it ensures that the likelihood decreases quickly as a function of permutation rank.

\section*{Following the expansion dynamics of a single-atom wave packet}\label{sec:results}

We repeat the measurement above for initial wave packets in traps of different frequencies ranging from $\omega=2\pi\cdot 260(15)$\,kHz to $2\pi\cdot 760(30)$\,kHz, by varying $U_0$ from $7\%$ to $61\%$ of the maximum value. In a given run we typically produce 20 to 50 independent but identically prepared wave packets. For each preparation and expansion time $t$ we record 100 image pairs (1 and 2) from which we obtain an averaged histogram reflecting the probability density. In Fig.\:\ref{fig:mosaic}, we show experimentally measured probability densities for each of these wave packets at different times of their dynamics, alongside the extracted evolution of their Gaussian width $\sigma(t)$, which shows the expected linear behaviour in $t$ at long times (see Fig.\:\ref{fig:mosaic}c). Indeed, for a wave packet prepared with an average vibrational number $\langle n \rangle=(\left\langle n_x \right\rangle+\left\langle n_y\right\rangle)/2$, the Schr{\"o}dinger equation predicts that its width $\sigma(t)$ evolves as:
\begin{eqnarray}
\sigma^2(t)&=& \sigma^2 (0)+\frac{\Delta p^2(0)}{m^2} t^2\\
\label{eq:expand}
&=& \left(2\langle n \rangle+1\right)\frac{\hbar}{2m\omega}\left[1+\omega^2t^2\right].
\end{eqnarray} In the long time limit $\omega^2t^2\gg1$, this simplifies to $\sigma(t)= \sqrt{\frac{\hbar \omega}{2m}}\sqrt{2\left\langle n \right\rangle+1} \cdot t$. The slope of $\sigma(t)$ in this regime provides a direct measurement of $\langle n \rangle$. In Fig.\:\ref{fig:mosaic}d, we report the measured values  of 
$\langle n \rangle$ for all trap depths $U_0$, which all agree to better than 8\% with the overall average value of $\langle n \rangle=0.46(3)$, and are consistent with the expansion predicted by the Schr{\"o}dinger equation. For a thermal distribution of the vibrational state populations, this corresponds to $\sim50\,\%$ of atoms in the ground state $\ket{n_x=0,n_y=0}$.

These values include a small correction due to the non-instantaneous release of the wave packet. Indeed, the turn-off of the lattice power follows a decaying sigmoid with a characteristic time of $\sim250$\,ns. For the turn-off to be considered instantaneous, this time-scale should be negligible compared to the harmonic oscillator period $T=2\pi/\omega$. Otherwise, the trapped wave packet first experiences a continuous opening of the harmonic trap before release, resulting in an effective reduction of the kinetic energy, and a lower apparent $\left\langle n \right\rangle$ during the free expansion. We quantitatively account for this effect and predict downshifts of the average motional number $\delta n= [0.014(1), 0.030(1), 0.064(2), 0.108(3),0.173(4)]$ for increasing values of $U_0$ \cite{Note2}, with the largest impact for the two highest values of $U_0$, with $T=1.65(10)\,\mu$s and  $T=1.3(1)\,\mu$s. Correcting for this shift results in the actual values of $\left\langle n \right\rangle$. Our ability to quantitatively distinguish this subtle effect demonstrates the high level of control of our setup and the precision of our imaging protocol.

\begin{figure}[!ht]
    \includegraphics[width=0.5\textwidth]{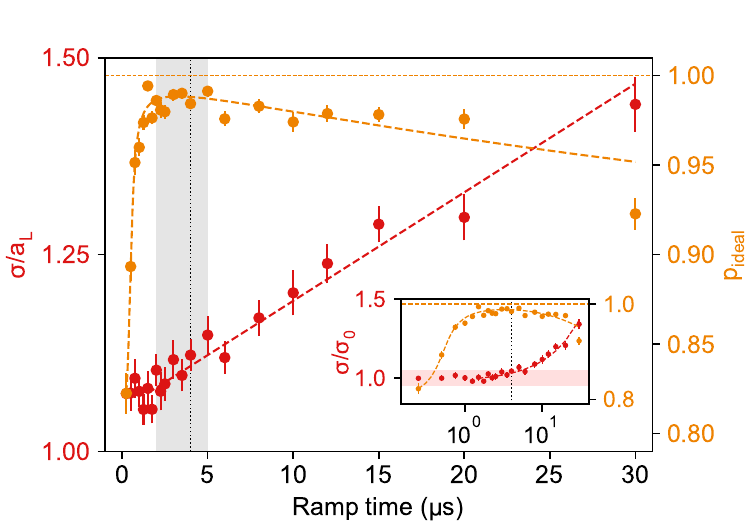}
    \caption{\textbf{Continuum pinning dynamics.} Effect of the lattice ramp time on the probability to be correctly pinned on the closest lattice site when projecting from continuous space ($p_\mathrm{ideal}$, yellow data points), and the width of the measured Gaussian probability density in units of lattice spacing ($\sigma/a_\mathrm{L}$, red data points) for $U_0/U_\mathrm{max} = 0.23$ and an expansion time of $4\,\mu$s. The vertical grey band displays the ramp time interval where $p_\mathrm{ideal}$ is maximal and the vertical dashed line shows the ramp time used for the data presented in Fig.\:\ref{fig:mosaic}. The yellow dotted line indicates $p_\mathrm{ideal}=1$, representing perfect pinning from continuum. The inset shows a semi-logarithmic plot of $p_\mathrm{ideal}$, and $\sigma$ normalized to the initial width, with the horizontal red band indicating a $\pm 5\%$ interval. Dashed lines represent guides to the eye for the respective data sets. Error bars give the standard error of the maximum likelihood estimates.}
    \label{fig:ramptime}
\end{figure}

\section*{Pinning from continuum}\label{sec:pinning}
 
The agreement between the measured and predicted wave packet dynamics also serves as a validation of our procedure for pinning atoms initially evolving in continuous space, which has never been addressed quantitatively. In most quantum gas microscope experiments, atoms are initially loaded in the lowest band of a shallow optical lattice, and the pinning is performed by ramping up the lattice power according to the adiabaticity criterion $\dot{\omega}/\omega^2 \ll1$.
By contrast, pinning atoms from the continuum leads to a more complex projection dynamics of the wavefunction. Although it is impossible to be strictly adiabatic -- due to the absence of an initial energy gap -- one can still mitigate the impact of diabatic processes. This requires setting a lower limit on the lattice ramp-on time $\tau$, as an instantaneous ramp-on would lead to a non-negligible projection onto higher bands, where atoms are less efficiently cooled. The final lattice frequency $\omega$ sets a suitable timescale for this requirement, i.e. $\omega\cdot\tau\gg 1$. On the other hand, a reliable measurement of the initial atom positions requires the lattice to be ramped fast enough to prevent any significant expansion of the wave packet over the ramp-up time $\tau$. Ideally, atoms should be projected to the nearest site upon pinning, which sets the condition that $v\cdot\tau \lesssim a_{\rm L}$, where $v$ is a characteristic velocity of the system. The lattice ramp time is therefore constrained by the double inequality $\omega^{-1} \ll \tau \lesssim a_{\rm L}/v$.

To quantitatively address this question, we performed the same type of wave packet expansion measurement, varying the ramp up time of the lattice $\tau$, for a fixed expansion time $t$. As figures of merit for the reliability of the pinning, we extract the Gaussian width $\sigma$, and the probability $p_{\rm ideal}$ to be pinned to the closest lattice site for different ramp up times. The probability $p_{\rm ideal}$ is directly provided by our maximum likelihood analysis \cite{Note2}. In Fig.\:\ref{fig:ramptime}, we present a typical optimization measurement showing that $p_{\rm ideal}$ first increases rapidly to reach a maximum value of $99\%$ for $\tau=2-5\,\mu$s and starts to slowly decrease for longer times. On the other hand, the Gaussian width $\sigma$ is essentially constant up to $\tau=5\,\mu$s and then follows a linear increase. In terms of the previously discussed time contraints, $\sigma$ is expected to grow with $\tau$ and hence serves to identify the upper time limit, while keeping $p_{\rm ideal}\approx1$ sets the lower time limit. We are thus in a favorable situation where we have a range of values for $\tau$, where $p_{\rm ideal}$ is maximal while $\sigma$ remains close to its $\tau=0$ value. This is quantitatively consistent with the double inequality above, which in this specific case yields $0.16\,\mathrm{\mu s} \ll \tau \lesssim 4\,\mathrm{\mu s}$. For all the data presented in the previous section, we have used a fixed ramp time $\tau=4\,\mu$s for the sake of consistency, with a negligible cost on the precision of $\sigma$ for the highest trap depths.

This study establishes a general requirement on the pinning timescale for the projection of an extended wavefunction from continuous space, which is constrained by two limits. While the lower time limit will remain fixed for most experiments, the upper limit depends on the typical particle velocity. For instance, for a 2D degenerate Fermi gas with an inter-particle spacing of a few lattice sites, the typical Fermi velocities are $\sim 0.02\,$m/s, an order of magnitude smaller than the particle velocities probed here. The upper ramp-up time limit could therefore be relaxed even further in that case, which implies that our imaging method is directly applicable to such a system.

\section*{Outlook}\label{sec:conclusion}

In this work, we observed the textbook expansion dynamics of a one-atom Gaussian wave packet using a new protocol for single-atom-resolved imaging in continuous space. The excellent agreement obtained here with the scaling dynamics predicted by the Schr{\"o}dinger equation constitutes a crucial benchmark for the reliability of our imaging method. Our work represents a milestone towards applying quantum gas microscopy to continuous-space many-body systems in the near future.

Alongside recent advances in single-atom imaging of few particle systems \cite{holten2021,holten2022,brandstetter2023}, our approach opens radically new possibilities for exploring correlated bulk quantum gases at the microscopic level, for instance by allowing direct access to spatially-resolved correlation functions. In combination with currently available homogeneous potentials \cite{gaunt2013, chomaz2015,mukherjee2017,navon2021}, our imaging technique will facilitate the search for exotic phases of matter that have been elusive so far, such as the Fulde-Ferrell-Larkin-Ovchinnikov superfluid \cite{fulde1964,larkin1965} or atomic quantum Hall fluids with fractional statistics \cite{cooper2008,ho2016,repellin2017}.

\section*{Acknowledgements}

We thank Shuwei Jin for instrumental early contributions to the planning and construction of the apparatus, Immanuel Bloch and Zoran Hadzibabic for insightful discussions, Dorian Gangloff, Sebastian Will and Martin Zwierlein for a critical reading of the manuscript. We are indebted to Antoine Heidmann for his crucial support as head of Laboratoire Kastler Brossel. This work has been supported by Agence Nationale de la Recherche (Grant No. ANR-21-CE30-0021), CNRS (Tremplin@INP 2020), and R{\'e}gion Ile-de-France in the framework of DIM SIRTEQ and DIM QuanTiP. We acknowledge support from the European Research Council in the construction phase of the apparatus.

\newpage

\renewcommand\thefigure{S\arabic{figure}}
\setcounter{figure}{0} 

\renewcommand\theequation{S\arabic{equation}}
\setcounter{equation}{0} 

\noindent

\section*{Supplementary Materials}\label{sec:Methods}

\subsection*{Experimental sequence}

In this section we provide additional details on the preparation and measurement protocol of the single-atom wave packets. After a laser cooling sequence described in \cite{jin2024}, we load a balanced mixture of $\sim 4\times 10^6$ $^6$Li atoms in the two lowest hyperfine states of the electronic ground state manifold in the Paschen-Back regime -- denoted as $|1\rangle$ and $|2\rangle$ -- in a cylindrically symmetric optical dipole trap (ODT) created by a 156\,W, 1070\,nm ytterbium fiber laser. The ODT is perpendicularly crossed with a highly oblate 1064\,nm laser beam, a light sheet with a $1/e^2$-waist of $w_x = 80\,\mathrm{\mu m}$, $w_z = 6\,\mathrm{\mu m}$ in the horizontal and vertical direction respectively, and a maximum power of $10\,\mathrm{W}$. We initiate evaporation by ramping down the power of both beams to $2.5\,\%$ of their initial value, while maintaining a magnetic field of 832\,G -- corresponding to the center of the broad Feshbach resonance between states $|1\rangle$ and $|2\rangle$. The cylindrical ODT is subsequently ramped off, fully loading the atoms in the potential of the light sheet, and the magnetic field is ramped to 0\,G, effectively removing particle interactions.

In order to create several independently expanding single-particle wave packets, we reduce the atom number to $N \sim 20 - 50$ by briefly turning off the light sheet for $3\,\mathrm{ms}$ before ramping it back to its maximum power and turning on the triangular lattice. The latter is created by the self-interference of a single $1064\,$nm laser beam ($w_x \simeq w_y \simeq 80\,\mathrm{\mu m}$, $P_{\mathrm{max}} = 30\,\mathrm{W}$), crossing at $120^{\circ}$ angles, as described in the main text. The lattice frequencies differ due to a slight anisotropy, leading to $\omega_x \simeq 2\pi \times 1020(50)\,\mathrm{kHz}$ and $\omega_y \simeq 2\pi \times 930(50)\,\mathrm{kHz}$. The average frequency $\omega = (\omega_x+\omega_y)/2$ is independently calibrated using intensity modulation and Raman spectroscopy, while the anisotropy is quantified through the method described in Ref.\,\cite{jin2024}. In the following, we ignore the anisotropy for simplicity and set $\omega_x \simeq \omega_y \simeq \omega$, although the exact frequencies are used for data analysis. 

At full power, tunneling is strongly suppressed and atoms are locally confined to a single lattice site in the $xy$-plane.  The light sheet ensures vertical confinement with a maximum trapping frequency of $\omega_z = 2\pi \times 50(5)\,\mathrm{kHz}$. We then initiate Raman sideband cooling (RSC) simultaneously with the triangular lattice ramp-on. An initial cooling stage of $3\,\mathrm{s}$ brings the individual atoms close to the ground state of the individual lattice sites. Further details on the RSC are provided in the next section.

After the first RSC stage we record a single-atom resolved fluorescence image with an exposure time of $600\,\mathrm{ms}$. The imaging system consists of a microscope objective, followed by a 1500\,mm focal length lens and an EMCCD camera. The objective has a numerical aperture of 0.56, an effective focal length of 27\,mm and a working distance of 16\,mm, resulting in a measured magnification of $M=59.7(6)$. The vertical confinement provided by the light sheet ensures that all atoms lie within the 2-$\mathrm{\mu m}$ depth of field of the microscope objective. During exposure, we collect around $\sim 800$ photons per atom on the camera, 6.5\,\% of the total number of photons emitted.

This first image allows us to register the initial positions of the individual atoms in the dilute sample. We then switch off RSC beams and adiabatically decrease the lattice and light sheet powers to a fraction of their maximum value (ranging from $7\,\%$ to $61\,\%$ of the maximum for the lattice, and a fixed $10\,\%$ for the light sheet), before turning the lattice off completely and letting each wave packet expand for a certain time $t$. During expansion, we keep the light sheet power at 10\% of its maximum value to maintain vertical confinement, while the effects of the in-plane light sheet potential are negligible for the considered expansion times. Finally, the lattice and light sheet are again ramped up to high power, after which we turn the RSC beams back on. We then expose the camera for $600\,\mathrm{ms}$ a second time, and obtain another single-atom image with different atom positions due to the expansion of each wave packet.

\subsection*{Raman sideband cooling and single atom imaging}

The atomic trapping potential of the individual lattice sites at the bottom of the wells is accurately modeled by two-dimensional harmonic oscillators with an energy spacing given by the lattice trapping frequency $\omega$. We hence denote their motional quantum states as $\ket{n_x, n_y}$, corresponding to an energy $E = (n_x+n_y+1)\,\hbar \omega$. Raman sideband cooling is used to bring the trapped atoms close to the motional ground state $\ket{n_x=0,n_y=0}$ of each lattice site during the initial preparation as well as to provide fluorescence to obtain single atom images.
It achieves this by coupling a change of the atomic internal state with a change of the motional state through a two-photon Raman transition \cite{heinzen1990,kerman2000}.

At zero magnetic field, the electronic ground state manifold of $^6$Li consists of the $|g_1\rangle = |2^2 S_{1/2}, F=1/2\rangle$ and $|g_2\rangle = |2^2 S_{1/2}, F=3/2\rangle$ hyperfine levels, with $F$ the hyperfine angular momentum quantum number. When the atoms are trapped in the sites of the two-dimensional lattice a full description of their eigenstates must include the motional quantum numbers: $\ket{F,n_x,n_y} = \ket{2^2 S_{1/2},F}\ket{n_x,n_y}$. During RSC two Raman beams, R$_1$ and R$_2$, drive a two-photon transition between $\ket{F=3/2,n_x,n_y}$ and either $\ket{F=1/2,n_x-1,n_y}$ or $\ket{F=1/2,n_x,n_y-1}$. The beam geometry, represented in Fig.\:\ref{fig:general_procedure}c of the main text, with R$_1$ along $y$ and R$_2$ along $x-z$, is chosen such that the differential wave vector $\mathbf{\delta k}=\mathbf{k_1-k_2}$ has a non-zero projection along all three spatial directions, ensuring momentum transfer along each, and thus coupling to different motional states. The two-photon Rabi frequencies $\Omega_i$ with $i \in \{x,y,z\}$, are directly proportional to the respective momentum-displacement operators: $\Omega_i \propto \left\langle{n_i-1}\right|e^{i \mathbf{\delta k}\cdot \mathbf{r}}\ket{n_i}$.
Dissipation is introduced through the addition of a repumper (RP) tuned to the atomic $D_1$ transition ($2^2 S_{1/2} \rightarrow 2^2 P_{1/2}$), exciting the atoms in the $\ket{2^2 S_{1/2}, F = 1/2}$ to the $\ket{2^2 P_{1/2}, F' = 1/2}$ internal state, followed by a spontaneous decay back into the electronic ground state. For the cooling to be efficient, the repumping process should preferentially conserve the motional quantum numbers. This propensity is quantified by the Lamb-Dicke parameter $\eta_i = \sqrt{\hbar/(2m\omega_i)}\delta k_i$ and holds when $\eta_i \sqrt{n_i}\ll 1$. In this Lamb-Dicke regime, each cooling cycle reduces the motional quantum number by one.

In the combined $xy$-lattice and light sheet, we obtain the following Lamb-Dicke factors $\eta_x = 0.19(1)$, $\eta_y = 0.28(1)$ and $\eta_z = 0.86(4)$ at maximal power. Although we do not strictly reach the Lamb-Dicke limit in the vertical direction, this does not prevent us from pinning the atoms in the $xy$-plane, as we have experimentally verified that RSC keeps the atoms trapped in a lattice site for several seconds, orders of magnitude longer than the observed lifetime in the absence of cooling.

In order to minimize parasitic one-photon transitions arising from off-resonant processes and to favor two-photon transitions, R$_1$ and R$_2$ are both blue-detuned with a one-photon detuning of about $\Delta \sim 2\pi\times3\,\mathrm{GHz}$ with respect to the $D_1$ transition. Additionally, the two-photon detuning $\delta=2\pi\times1.5\,\mathrm{MHz}$ with respect to the $F=1/2 \to F=3/2$ hyperfine transition corresponds to $\sim 1.5 \omega$, which provides a good balance between cooling power and fluorescence. We attribute the necessity to be red-detuned from the $\ket{n} \to \ket{n-1}$ transition to the presence of power broadening, which increases the probability of an off-resonant $\ket{n} \to \ket{n}$ carrier transition. Finally, the repumper frequency is centered on the $F=1/2 \to F'=1/2$ transition of the $D_1$ line, as this excited state has a high probability to decay back to $F=3/2$, maximizing the repumping efficiency.

The Raman beam characteristics are presented in Table\,\ref{table:Raman_parameters}. In absence of the lattice we measure a bare two-photon Rabi frequency $\Omega = 2\pi \times 800(50)\,\mathrm{kHz}$. At full lattice power, the motional-state coupling alters the Rabi frequencies by the respective Lamb-Dicke factors leading to values of $\Tilde{\Omega}_x = \eta_x \Omega = 2\pi \times 150(15)\,\mathrm{kHz}$ and $\Tilde{\Omega}_y = \eta_y \Omega = 2\pi \times 220(20)\,\mathrm{kHz}$, corresponding to the $n=1 \to n=0$ transition. The repumping rate $F=1/2 \to F'=1/2 \to F=3/2$ was experimentally determined to be on the order of $10^6\,$s$^{-1}$.

\begin{table}[!t]
\renewcommand{\arraystretch}{1.3}
\caption{Raman beam parameters. Intensities are provided in units of the $D_1$ saturation intensity ($7.59\,\mathrm{mW/cm^2}$).}
\begin{tabular}{l c c c}
\hline \hline
\textbf{Beam} & \textbf{R$_\mathbf{1}$} & \textbf{R$_\mathbf{2}$} & \textbf{RP} \\ \hline \hline
Waist ($\mathbf{\mathrm{\mu m}}$)             & $120$  & $120$ & $5000$\\
Power (mW)                                    & $4.0$  & $0.4$ & $2.0$ \\
Intensity (${I_{\mathrm{sat}}^\mathrm{D_1}})$ & $2300$ & $230$ & $0.7$\\
Propagation axis & $\mathbf{e_y}$ & $\frac{\mathbf{e_x} - \mathbf{e_z}}{\sqrt{2}}$ & $\mathbf{e_x}$\\
Polarization & $\mathbf{e_z}$ & $\mathbf{e_y}$ & $\mathbf{e_z}$ \\ \hline \hline
\end{tabular}%

\label{table:Raman_parameters}
\end{table}

\subsection*{Numerical Simulation of the Raman Cooling Process}

Using the experimental RSC and lattice parameters, we determine the RSC efficiency and estimate the average motional quantum number $\langle n\rangle$ of the atoms in the steady state by numerically solving the Lindblad master equation that describes the time evolution of the density matrix $\rho$ throughout the cooling process:

\begin{equation}
\frac{d \rho}{dt} = - \frac{i}{\hbar} \left[H,\rho\right] + \sum_{\mu} \left(L_{\mu} \rho L_{\mu}^{\dagger} - \frac{1}{2} L_{\mu}^{\dagger} L_{\mu} \rho - \frac{1}{2} \rho L_{\mu}^{\dagger} L_{\mu} \right).
\label{eq:Lindblad_equation}
\end{equation}

The first term on the right-hand side of Eq.\:(\ref{eq:Lindblad_equation}) corresponds to the Hamiltonian ($H$) evolution of the system while the second part captures the decohering and dissipative processes through a set of jump operators $L_{\mu}$, where $\mu$ indexes the relevant transitions. For instance, spontaneous emission from an excited state $\ket{e}$ to the ground state $\ket{g}$ is represented by the operator $L = \sqrt{\Gamma} |g \rangle \langle e |$, where $\Gamma$ is the linewidth of the excited state.

The relevant Hilbert space is spanned by the product of the electronic and motional states of the atom. At zero magnetic field we can simplify the electronic structure to the two ground states $\ket{g_1}$ and $\ket{g_2}$, as well as two excited states; $\ket{e_1}$, which represents the $\ket{2^2 P_{1/2},F=1/2}$ state addressed by the repumper beam, and $\ket{e_2}$, which encompasses all other excited states from the $D_1$ and $D_2$ transitions. The motional spectrum consists of the two dimensional harmonic oscillator eigenstates $\ket{n_x,n_y}$ with associated energy $(n_x+n_y+1) \hbar \omega$. We include a hard cutoff on the maximal number of excitations $n_x + n_y \leq N_{\mathrm{cutoff}}=10$, which we verify to be sufficient for convergence of all relevant quantities.

\begin{figure}[!b]
    \centering
    \hspace*{0.5em}
    \includegraphics[width=7.6368cm]{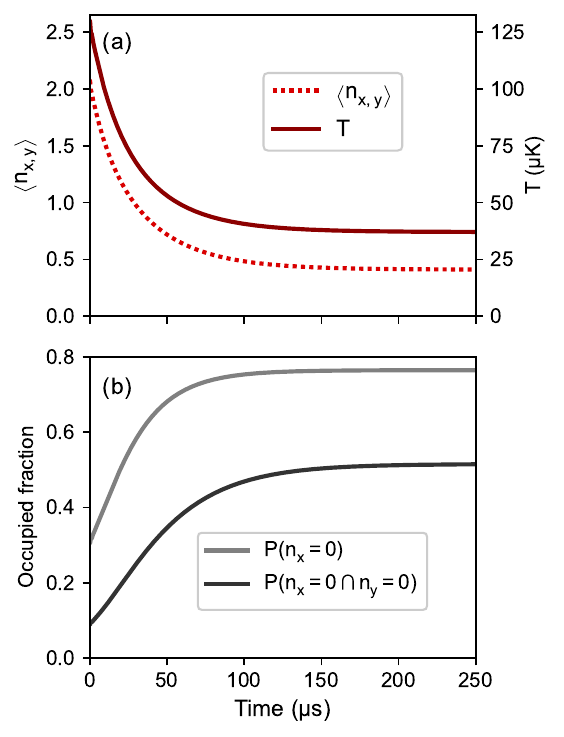}
    \caption{\textbf{Simulated time-evolution of the motional degrees of freedom during RSC.} Atoms start in a thermal superposition of harmonic oscillator states with $\left\langle n_x \right\rangle = \left\langle n_y \right\rangle = 2$. \textbf{(a)} Evolution of the average quantum number $\left\langle n_{x,y}\right\rangle$ and temperature over time. Temperature is obtained by fitting the $(n_x,n_y)$ populations to a Boltzmann distribution. \textbf{(b)} Evolution of the fraction of atoms with $n_x=0$ (grey line), and the fraction of atoms in the absolute ground state $\ket{n_x=0,n_y=0}$ (black line).}
    \label{fig:RSC_plot}
\end{figure}

\begin{figure*}[!ht]
    \centering
    \includegraphics[width=16.8cm]{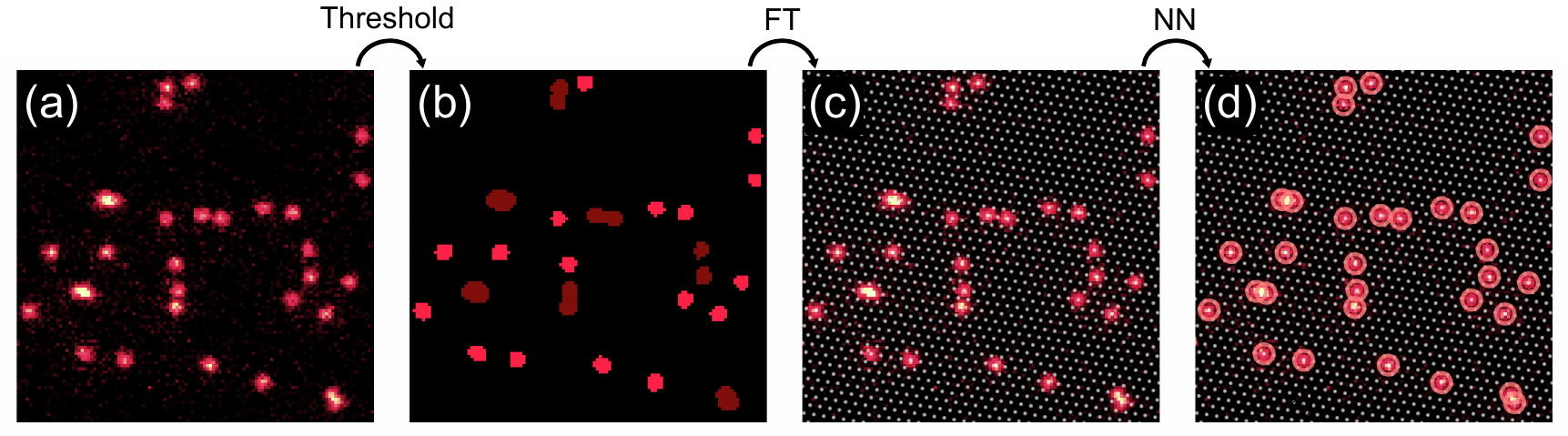}
    \caption{\textbf{Image reconstruction procedure}. \textbf{(a)} Raw experimental image. \textbf{(b)} A simple threshold is applied to distinguish the signal from the background. Contiguous bright pixels are grouped together in clusters, and only those that are well-separated and have a size comparable to the experimental point-spread function (light red clusters) are used for lattice reconstruction. Others (dark red clusters) are not used for lattice reconstruction. \textbf{(c)} Lattice site positions (white dots) are found through a Fourier transform (FT) on the center-of-mass position of each of selected clusters. \textbf{(d)} A small region of interest of $9\times9$ pixels around each lattice site is extracted from the image and sent through the neural network (NN) described in Fig.\:\ref{fig:neural_network}, which was trained to distinguish between occupied and unoccupied sites. Identified atoms are indicated by red circles.}
    \label{fig:image_analysis}
\end{figure*}

The Hamiltonian includes the two-photon transition between $\ket{g_1}$ and $\ket{g_2}$ with a Rabi frequency $\Omega$, as well as the repumper beam coupling $\ket{g_1}$ to the excited state $\ket{e_1}$ with a Rabi frequency $\Omega_p$ (both determined through experimental calibration). Spontaneous emission is taken into account via the jump operators $L_{i,j} = \sqrt{\Gamma_{i,j}}  |g_j \rangle \langle e_i |$, with decay rates $\Gamma_{i,j}$ obtained from the $^6 \mathrm{Li}$ electronic structure. Off-resonance absorption events from Raman beams are included through an effective jump operator $L_{\mathrm{heating}} = \sqrt{\gamma_{\mathrm{heating}}} |e_2 \rangle \langle g_2 |$, where $\gamma_{\mathrm{heating}}$ is estimated from an independent experimental calibration.

We get the transition coefficients between any two harmonic oscillator states by multiplying the bare coupling constants defined above ($\Omega$, $\Omega_p$, $L_{i,j}$ and $L_{\mathrm{heating}}$) by the factor $\left\langle{n'_x,n'_y}\right| e^{i \mathbf{\delta k} \cdot \mathbf{r}} \ket{n_x,n_y}$, which only depends on the Raman beams geometry and are computed through the corresponding integral over the eigenstates of the harmonic oscillator. To obtain the equilibrium populations, we initialize the density matrix to an arbitrary configuration $\rho_0$ and let it evolve according to Eq.\:(\ref{eq:Lindblad_equation}) until a steady-state is reached. A typical time evolution of motional degrees of freedom is shown in Fig.\:\ref{fig:RSC_plot}. We also obtain an uncertainty estimate on $\left\langle n_{x,y} \right\rangle$ by repeating the simulation with model parameters randomly drawn from a distribution reflecting their experimental uncertainty. This results in an average of $\left\langle n_{x,y} \right\rangle = 0.4(1)$ across all realizations, in good agreement with the experimentally determined values.

\subsection*{Image analysis} \label{sec:img}
We extract atomic positions from the fluorescence images obtained with our microscope through a two-step analysis. For each image, we first reconstruct the structure of the pinning lattice to identify the position of the individual lattice sites. We then determine the occupancy of each previously identified lattice sites using a high-accuracy neural network.

Fig.\:\ref{fig:image_analysis} shows an example of the image analysis procedure on part of a single-atom image. As an initial step, lattice reconstruction is performed by identifying clearly resolvable atoms and performing a Fourier transform on their positions. We first apply a simple threshold on the pixel values, separating the signal (active) from the background (inactive). Active pixels are subsequently joined into contiguous groups of pixels connected to each other on at least one side. To remove effects of background noise, the lattice reconstruction only considers pixel groups with a size compatible with the experimental point-spread function (light red clusters in Fig.\:\ref{fig:image_analysis}b, with the dark red clusters corresponding to those not matching the expected size). We then apply a Fourier transform on the center-of-mass positions of these contiguous active pixel groups. This gives us the reciprocal lattice vectors, from which we reconstruct the positions of all individual lattice sites on each image (Fig.\:\ref{fig:image_analysis}c).

\begin{figure}[!b]
    \centering
    \vspace*{-1.5em}
    \includegraphics[width=8.315620656cm]{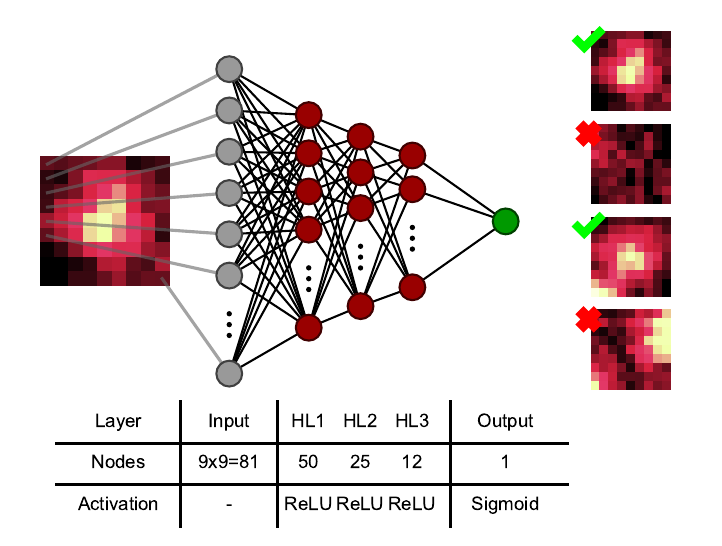}
    \caption{\textbf{Neural network architecture for atom recognition.} For each lattice site, we select a $9\times9$ pixels region of interest (ROI), which serves as the input layer (81 nodes). The network itself consists of this input layer, 3 fully connected hidden layers (HL1, HL2 and HL3) -- with 50, 25 and 12 nodes, respectively, and a ReLU activation function -- and a single output node with a sigmoid activation. The network is trained to recognize an atom located at the center of the ROI, and will ignore neighboring occupied sites, as seen from the negative result on the fourth image.}
    \label{fig:neural_network}
\end{figure}

We determine the occupancy of each reconstructed lattice site using a neural network trained to recognize the atomic fluorescence pattern. We take a $9\times 9$ pixel region centered on each site and pass the resulting array of pixel values to the neural network, which consists of three fully connected hidden layers and an output layer (see Fig.\:\ref{fig:neural_network}). To train the network, we simulated single-atom images by randomly placing atoms on the triangular lattice structure, creating an atomic fluorescence pattern at each occupied site using the experimentally determined PSF subject to Poissonian fluctuations, and adding overall Poissonian background noise to account for unfiltered background light. To increase robustness of the pattern recognition, we deliberately chose the simulated signal-to-noise ratio to be a factor $2$ to $3$ worse than in typical imaging conditions, and the atomic density to be $5$ to $100$ times larger than for our experimental images. Even in these highly degraded conditions, the neural network differentiates occupied and unoccupied sites with $99.5\,\%$ accuracy.

To quantify the percentage of atoms that remain in their pinned lattice site through RSC, we take two successive single atom images (without release) by exposing the camera for $600\,\mathrm{ms}$, followed by $400\,\mathrm{ms}$ wait time and a second exposure of $600\,\mathrm{ms}$, all while maintaining the lattice and Raman beams. We find that $99.0(3)\,\%$ of atoms identified on the first picture are found on the same exact lattice site on the second picture, with a hopping fraction of $0.8(2)\,\%$ and a loss fraction of $0.2(1)\,\%$ (estimated from the analysis of $\sim 1500$ pairs of single atom images).

\vspace*{-1em}
\subsection*{Maximum likelihood estimation and assignment problem}

\begin{figure}[!t]
    \centering
    \includegraphics[width=7.3cm]{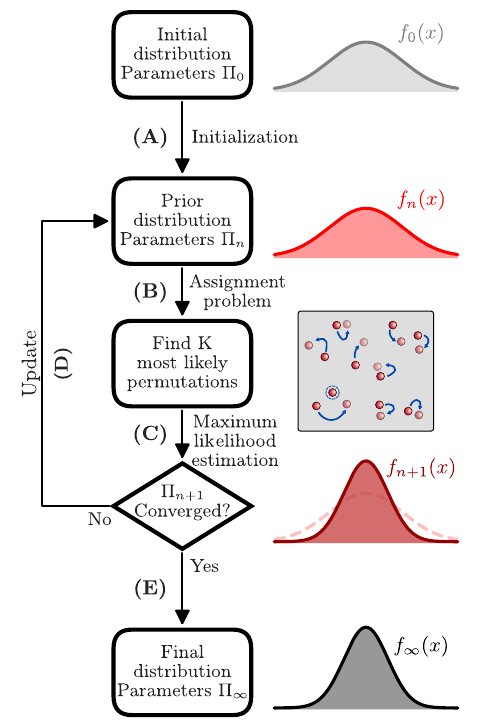}
    \caption{\textbf{Maximum likelihood estimation algorithm.}  We obtain the probability distribution of the relative position of atoms after lattice release and recapture, using the iterative procedure shown in the flowchart. After initialization (A), the prior distribution $p_n$ is used in (B) to compute the $K$ most likely permutations of atoms. These are used to update the prior distribution with a maximum likelihood estimation for the parameters (C). The obtained posterior distribution is used as a new prior (D). These three steps are repeated until convergence is reached, leading to the final parameters (E).}
    \label{fig:flowchart_MLE}
\end{figure}

Reconstructing the probability distribution associated with the expanding wave packet, as shown in Fig.\:\ref{fig:mosaic} of the main text, requires matching atoms in the first image with their corresponding position on the second image. While we are interested in single-particle behavior, we load multiple atoms in different lattice sites (typically $20-50$). This allows us to obtain multiple independent realizations in a single sequence, reducing the measurement time and decreasing the sensitivity to various changes in experimental conditions, with the added benefit of making the lattice reconstruction (as described in the Image Analysis section) more robust. In assigning the atoms detected after expansion to their initial position, all possible permutations are in principle allowed due to the indistinguishability of identical particles. In order to quantitatively determine the relevance of each permutation, we follow a self-consistent approach based on a maximum likelihood estimation summarized in Fig.\:\ref{fig:flowchart_MLE} and presented in the following.

This maximum likelihood analysis consists in an iterative optimization of the parameters of the model described below, which represents the probability distribution for the distance traveled by the atoms between the two images. Because we pin from the continuum, we model the distance traveled by the atoms between the first and second image with a multi-modal distribution, written as the sum of three contributions, corresponding to the following scenarios for the atom:

\begin{itemize}
 \item Being recaptured on the nearest lattice site upon pinning, occurring with probability $p_{\mathrm{ideal}}$. The associated probability density is a 2D Gaussian distribution with standard deviations $\sigma_x$, $\sigma_y$, i.e., the expected density distribution of the expanding wave packet.
 \item Being projected onto a high lattice band, leading to inefficient RSC and resulting in hovering above the lattice, occurring with probability $p_{\mathrm{hover}}$. In this case the recapture occurs stochastically and the probability density for the corresponding trajectory is a decaying exponential of characteristic size $L\gg \sigma_x,\sigma_y$.
 \item Being lost completely with a probability $p_{\mathrm{loss}}=1-p_{\mathrm{ideal}}-p_{\mathrm{hover}}$.
\end{itemize}

The total distribution is therefore:

\begin{equation}
f(x,y) = \frac{p_{\mathrm{ideal}}}{2 \pi \sigma_x \sigma_y} e^{-\frac{x^2}{2\sigma_x^2}-\frac{y^2}{2\sigma_y^2}}+ \frac{p_{\mathrm{hover}}}{2 \pi L \sqrt{x^2+y^2}} e^{-\frac{\sqrt{x^2+y^2}}{L}}.
\end{equation}

The model is thus described by five parameters $\Pi = \{p_{\mathrm{ideal}},\sigma_x,\sigma_y,p_{\mathrm{hover}},L\}$.  At the beginning of the optimization, we initialize the model parameters to realistic values $\Pi_0$ (step A in Fig.\:\ref{fig:flowchart_MLE}), which are improved through an iterative process described below.

The parameter set at the start of the $n^{\mathrm{th}}$ iteration is denoted $\Pi_n$, with an associated probability distribution $f_n(\mathbf{r})$. For the $i^{\mathrm{th}}$ experimental run of a given lattice power and release time, $f_n(\mathbf{r})$ is used as a prior distribution to compute the log-likelihood matrix $M^{(n,i)}$ for atoms to go from any of the initial position to any of the final positions:
\begin{equation}
M^{(n,i)}_{\alpha,\beta} = \mathrm{log}\left(f_n\left(\mathbf{r}=\mathbf{r}_{1,i,\alpha}-\mathbf{r}_{2,i,\beta}\right)\right),
\label{eq:log_likelihood_matrix}
\end{equation}
where $\mathbf{r}_{1,i,\alpha}$ and $\mathbf{r}_{2,i,\beta}$ are the positions corresponding to atom $\alpha$ on the first image, and atom $\beta$ on the second image of experimental realization $i$, respectively.

\begin{figure}[!ht]
    \centering
    \includegraphics[width=8.06106cm]{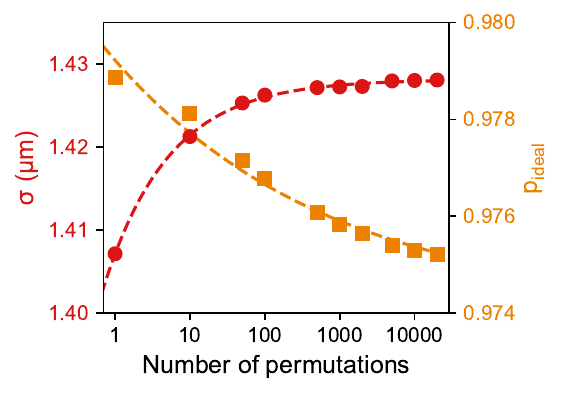}
    \caption{\textbf{Convergence of model parameters.} Evolution of $\sigma = \frac{\sigma_x+\sigma_y}{2}$ (red circles) and $p_{\mathrm{ideal}}$ (yellow squares) as a function of the largest permutation rank $K$ taken into account for the maximum likelihood estimation in equation (\ref{eq:total_likelihood}), for $U_0 = 0.23\, U_{\mathrm{max}}$ and $t=8\,\mathrm{\mu s}$. Dashed lines are guides to the eye for the respective parameters.}
    \label{fig:convergence_plot}
\end{figure}

Finding the most likely pairing of atoms is then equivalent to finding the permutation of matrix columns maximizing the trace. This problem, known as the linear assignment problem \cite{burkard1999}, can be solved in polynomial time $O(N^3)$, where $N$ is the matrix size (compared to $O(N!)$ for the brute force method consisting in testing each possible permutation). The algorithm can be generalized to find the $K$ most likely configurations with a $O(KN^4)$ time complexity (step B). For data analysis performed in this article, we used $K=20000$ permutations, at which point the relative likelihood has decreased to a typical value of $10^{-8}$ - $10^{-16}$ (see Fig.\:\ref{fig:comparison_likelihood}). We indeed reach convergence of model parameters at such values of $K$, as shown in Fig.\:\ref{fig:convergence_plot}. The likelihood $\mathcal{L}_{n,i}$ for image $i$ is then taken as the sum of the likelihood of these $K$ permutations:

\begin{figure}[!b]
    \centering
    \includegraphics[width=8.145914112cm]{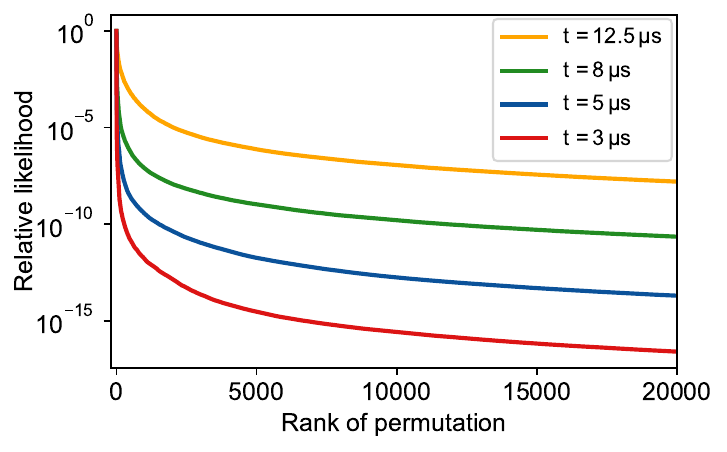}
    \caption{\textbf{Typical relative likelihood estimates for different release times.} Evolution of the relative likelihood for the $20000$ most likely permutations for $U_0 = 0.38\,U_{\mathrm{max}}$ and $t=3\,\mathrm{\mu s}$ (red line), $t=5\,\mathrm{\mu s}$ (blue line), $t=8\,\mathrm{\mu s}$ (green line) and $t=12.5\,\mathrm{\mu s}$ (yellow line), and a total atom number of $N\simeq 25$. For larger release times $t$ the likelihood decreases less rapidly due to the atoms traveling further.}
    \label{fig:comparison_likelihood}
\end{figure}

\begin{equation}
\mathcal{L}_{n,i} = \sum_{k=1}^{K} \mathrm{exp}\left(\sum_{\alpha=1}^{N} M^{(n,i)}_{\alpha, \mathfrak{S}_{n,i,k}(\alpha)}\right),
\label{eq:total_likelihood}
\end{equation}
where $\mathfrak{S}_{n,i,k} : \left[1 ... N \right] \to \left[1 ... N \right]$ is the $k^{\mathrm{th}}$ most likely permutation of atoms for experimental run $i$, calculated during the $n^{\mathrm{th}}$ iteration of the maximum likelihood routine.

Finally, model parameters are optimized to maximize the total log-likelihood $\mathrm{log}(\mathcal{L}_{n})= \sum_i \mathrm{log}(\mathcal{L}_{n,i})$ across all experimental realizations (step C), resulting in parameter set $\Pi_{n+1}$. The posterior distribution obtained from the maximum likelihood estimator is then used as the prior distribution $f_{n+1}$ for iteration $n+1$ (step D). Multiple iterations are required for self-consistency, because computing the $K$ most likely configurations relies itself on the distribution that we are trying to estimate. Steps (B), (C) and (D) are repeated until convergence of the model parameters, at which point we obtain the final probability distribution $f_{\infty}$ (step E). This distribution is used to construct the various histograms in Fig.\:\ref{fig:mosaic}a of the main text by computing the $K$ most likely permutations for each image pair and averaging them based on their relative likelihood. Finally a linear fit of $\sigma_x$, $\sigma_y$ at different release times $t$ is used to obtain the average quantum number $n$ of the wave packet (Fig.\:\ref{fig:mosaic}b and \ref{fig:mosaic}c of the main text).

\subsection*{Impact of non-instantaneity of the trap release}

In this section, we describe how we correct the extracted average motional quantum numbers shown in Fig.\:\ref{fig:mosaic}d for the finite duration of the lattice ramp-down. As the lattice power is controlled using an acousto-optic modulator, this switch-off time is limited to $t_{\mathrm{off}} = 250\,\mathrm{ns}$. For the momentum distribution after release to reflect the one before release, $t_{\mathrm{off}}$ must be short with respect to the characteristic lattice timescale $1/\omega$, i.e., $\omega t_{\mathrm{off}} \ll 1$. This condition is not fulfilled for all of our data, especially for the highest value of $U_0$, where $\omega t_{\mathrm{off}}\simeq1.2$. This results in a difference between the average kinetic energy of the atoms before release, $\left\langle E_k \right\rangle_0$, and after release, $\left\langle E_k \right\rangle_f$, leading to the apparent decrease in $\langle n\rangle$. This effect is to a lesser extent present for lower values of $U_0$.

The resulting change in kinetic energy of the particles prior to release can be determined analytically, either through a full treatment of the time-dependent quantum harmonic oscillator \cite{lewis1969} or through application of the Ehrenfest theorem. Here, we follow the latter method. Given a Hamiltonian $H(t) = \frac{\hat{P}^2}{2M} + \frac{1}{2}M \omega^2(t) \hat{X}^2$, the Ehrenfest theorem leads to a system of three coupled differential equations:

\begin{align}
 \frac{d\left\langle \hat{P}^2 \right\rangle}{dt} &= - M \omega^2(t) \left\langle \hat{X}\hat{P} + \hat{P}\hat{X} \right\rangle \nonumber\\
 \frac{d\left\langle \hat{X}^2 \right\rangle}{dt} &= \frac{1}{M} \left\langle \hat{X}\hat{P} + \hat{P}\hat{X} \right\rangle \nonumber\\
 \frac{d\left\langle \hat{X}\hat{P} + \hat{P}\hat{X} \right\rangle}{dt} &= \frac{2}{M} \left\langle \hat{P}^2 \right\rangle - 2 M \omega^2(t) \left\langle \hat{X}^2 \right\rangle.
 \label{eq:motion}
\end{align}
Assuming we start in a thermal superposition of harmonic oscillator eigenstates before the lattice release, the initial conditions are:
\begin{equation}
\begin{cases}
 &\left\langle \frac{\hat{P}^2}{2M} \right\rangle_0 = \left\langle \frac{1}{2}M\omega_0^2\hat{X}^2 \right\rangle_0 = \frac{E_0}{2}\\
 &\left\langle \hat{X}\hat{P} + \hat{P}\hat{X} \right\rangle_0 = 0.
\end{cases}
\end{equation}

\begin{figure}[!b]
    \centering
    \includegraphics[width=7.63679448cm]{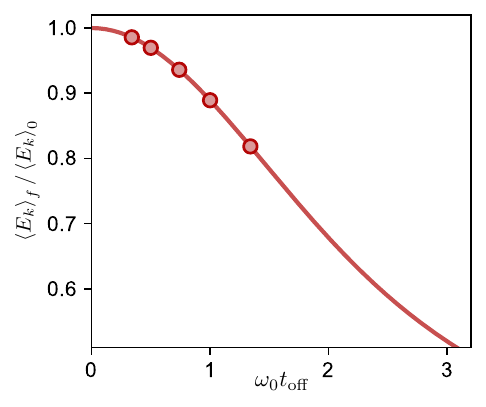}
    \caption{\textbf{Energy correction factor due to non-instantaneous lattice release.} Ratio of the final to initial kinetic energy of a quantum harmonic oscillator with a ramp-down profile $\omega(t)$ given by Eq.\:(\ref{eq:ramp}). Correction factors for the measurements presented in the main text are determined using a monitoring photodiode to extract the time profile of the lattice intensity and shown as red circles.}
    \label{fig:non_instantaneity}
\end{figure}

To determine the change in kinetic energy due to the non-instantaneous switch off, we use the $\omega(t)$ curve obtained from experimental photodiode signals at the different values of $U_0$. The ramp-down profile for each power is well-fitted by the following equation:
\begin{equation}
\label{eq:ramp}
    \omega^2(t) = \omega_0^2 \frac{1-\mathrm{erf}(2t/t_{\mathrm{off}})}{2},
\end{equation}
where $\mathrm{erf}$ is the error function, and $t_{\mathrm{off}} \simeq 250\,\mathrm{ns}$. The lattice frequency goes from $\omega_0$ for $t \to -\infty$ to $\omega = 0$ for $t \to \infty$.

By numerically integrating the system of equations (\ref{eq:motion}) we obtain the ratio of final to initial kinetic energy $\left\langle E_k \right\rangle_f/\left\langle E_k \right\rangle_0$, which only depends on the dimensionless parameter $\omega_0 t_{\mathrm{off}}$. Resulting values for this ratio at the various lattice frequencies and ramp times are shown in Fig.\:\ref{fig:non_instantaneity} and amount to $98.5(5)\,\%$, $97.0(5)\,\%$, $93.5(1.0)\,\%$, $89(1)\,\%$ and $82(1)\,\%$ for $U_0/U_\mathrm{max} = 0.07, 0.14, 0.23, 0.38$ and $0.61$, respectively. Reported uncertainties come from the experimental uncertainty in the ramp-off profile of the lattice intensity. 
Finally, we use these factors to correct the observed values of $\left\langle n \right\rangle$ for this non-instantaneous ramp-down, as shown in Fig.\:\ref{fig:mosaic}d.

\end{document}